%% file: our_paper.tex
\documentclass[10pt,letterpaper]{article}

\usepackage{cogsci}

\cogscifinalcopy %

\usepackage[
  style=apa,
  natbib=true,
  annotation=false,
]{biblatex}
\usepackage{float} %
\usepackage{amsmath}
\usepackage{amsfonts}
\usepackage{graphicx}
\usepackage{booktabs}
\usepackage{wrapfig}
\usepackage{url}
\usepackage[misc]{ifsym}
\newtheorem{theorem}{Theorem}
\newcommand{\err}[1]{\scriptsize $\pm$ #1}

\newcommand\nnfootnote[1]{%
  \begin{NoHyper}
  \renewcommand\thefootnote{}\footnote{#1}%
  \addtocounter{footnote}{-1}%
  \end{NoHyper}
}

\title{Modeling Decision-Making with Will for Cooperation in Social Dilemmas}

\author[1, 2]{\mbox{Yizhe Huang$^\textbf{*}$}}
\author[3]{\mbox{Bin Ling$^\textbf{*}$}}
\author[2, 1]{\mbox{Song-Chun Zhu}}
\author[2]{\mbox{Xue Feng~\textsuperscript{\Letter}}}
\affil[1]{School of Intelligence Science and Technology, Peking University}
\affil[2]{State Key Laboratory of General Artificial Intelligence, BIGAI}
\affil[3]{Law School, Peking University}

\begin{document}

\maketitle
\nnfootnote{$^\textbf{*}$Equal contribution.
$\textsuperscript{\Letter}$Corresponding author. The code is available at \url{https://github.com/j-alpha7/willed-agent-sd}.}

\begin{abstract}
Standard rational actor models often attribute cooperation failures in social dilemmas to insufficient incentives, overlooking the destabilizing effects of continuous utility maximization. To address this, we propose a framework of ``will" defined as a mechanism that persistently pursues goals while ignoring local cost-benefit fluctuations. We formalize the Willed Agents as potential minimizers, distinguishing them from cumulative utility maximization. Dynamical analysis of infinite population demonstrates that willed agents shrink the feasible state space, acting as boundary constraints that accelerate convergence in canonical social dilemmas. Through multi-agent simulations in a spatiotemporal Stag Hunt Game, we show that willed agents function as ``cooperation catalysts", enabling groups to surmount high-risk thresholds where purely utility maximization fails. We find that heterogeneous will strength promotes cooperation, and that agents who autonomously suspend rational re-evaluation can significantly outperform continuous optimizers. These findings suggest that successful cooperation relies on the cognitive capacity to strategically constrain calculation.

\end{abstract}
\vspace{-5pt}
\input{sections/introduction_3}

\section{Mathematical Framework}

We model the social environment as a \textbf{Markov Game with Will}, formalizing the distinction between agents driven by external rewards and Willed Agents driven by internal distance-minimizing constraints. The system is defined as the tuple $\mathcal{M} = \langle N, \mathcal{S}, \mathcal{A}, P, \mathbf{R}, \mathbf{G}, \mathbf{D}, T \rangle$:

\begin{itemize}
    \item $N = \{1, \dots, n\}$: The set of $n$ agents.
    \item $\mathcal{S}$: The joint state space representing the global system configuration.
    \item $\mathcal{A} = \mathcal{A}_1 \times \dots \times \mathcal{A}_n$: The joint action space, where $\mathcal{A}_i$ is the set of available actions for agent $i$.
    \item $P: \mathcal{S} \times \mathcal{A} \to \Delta(\mathcal{S})$: The state transition probability function, where $P(\mathbf{s}' \mid \mathbf{s}, \mathbf{a})$ is the probability of reaching $\mathbf{s}'$ given current state $\mathbf{s}$ and joint action $\mathbf{a}$.
    \item $\mathbf{R} = \{R_1, \dots, R_n\}$: Reward functions $R_i: \mathcal{S} \times \mathcal{A} \to \mathbb{R}$ for each agent $i$.
    \item $\mathbf{G} = \{G_1, \dots, G_n\}$: Target state sets $G_i \subset \mathcal{S}$ for agent $i$.
    \item $\mathbf{d} = \{d_1, \dots, d_n\}$: Pseudometrics $d_i: \mathcal{S} \times \mathcal{S} \to \mathbb{R}_{\ge 0}$ describes the distance between two states.
    \item $T \in \mathbb{N}$: Episode length.
\end{itemize}

\subsection{Will as Potential Minimization}

To formalize ``will,'' we define the Willed Agent by its direct minimization of distance to a goal, rather than by accumulated payoff. A Willed Agent $i$ is governed by a potential field $D_i$:

\begin{equation}
\label{eq:potential}
D_i(\mathbf{s}) = \inf_{\mathbf{g} \in G_i} d_i(\mathbf{s}, \mathbf{g}),
\end{equation}
where $d(\cdot, \cdot)$ is a pseudometric on the state manifold. 
Unlike rational agents that integrate scalar signals over time to \textbf{maximize cumulative extrinsic utility} $\sum_{t} R_i(s_t, \mathbf{a}_t)$, the Willed Agent operates via descent on the potential field to \textbf{minimize distance error} $D_i(\mathbf{s}_T)$, a bounded objective defined strictly by state space topology. Its decision rule greedily minimizes expected distance to $G_i$:

\begin{equation}
\pi_{will}(\mathbf{s}) = \operatorname*{arg\,min}_{a \in \mathcal{A}_i} \mathbb{E}_{\mathbf{s}' \sim P(\cdot \mid \mathbf{s}, a_i, \mathbf{a}_{-i})} [D_i(\mathbf{s}')].
\end{equation}

\subsection{The Potential Attractor}

A \textbf{Potential Attractor} emerges when a Willed Agent becomes physically ``pinned'' by environmental boundaries or interaction dynamics.

\textbf{Definition:} A state $\mathbf{s} \in \mathcal{S}$ is a Potential Attractor for agent $i$ if:
\begin{equation}
\label{potential_trap}
\forall a_i \in \mathcal{A}_i, \quad \mathbb{E}_{\mathbf{s}' \sim P(\cdot \mid \mathbf{s}, a_i, \mathbf{a}_{-i})} [D_i(\mathbf{s}')] \ge D_i(\mathbf{s}).
\end{equation}

Locally trapped in this potential well, an agent might fail to reach its exact goal ($D_i(\mathbf{s}) > 0$) if further progress requires coordinated joint actions. Inside a Potential Attractor, the Willed Agent acts as a \textbf{fixed boundary condition} in the phase space, resisting deviations from $G_i$.

\section{Dynamical Analysis of Population with Will}

Having formalized the Willed Agent as a potential minimizer, we analyze how these individual constraints scale to shape collective behavior. For tractability, we examine an infinite population ($N \to \infty$) making independent choices between two targets: $G_1$ (Cooperation) and $G_2$ (Defection).

\subsection{Population Structure and Payoffs}
\label{sec:structure}

Let the system state $x \in [0, 1]$ denote the total proportion of the population pursuing $G_1$. The population comprises two decision-making modes:

\begin{itemize}
    \item \textbf{Willed Mode:} Agents constrained by Potential Attractors (Eq.~\ref{potential_trap}). We define $n_1$ and $n_2$ as the fixed proportions rigidly committed to $G_1$ (Willed Cooperators) and $G_2$ (Willed Defectors), with $0 \leq n_1 + n_2 \leq 1$.
    \item \textbf{Rational Mode:} The remaining agents, $m = 1 - n_1 - n_2$, adapt strategies to maximize cumulative extrinsic reward.
\end{itemize}

Rewards $\mathbf{R}$ translate into frequency-dependent payoffs. Let $f_1(x)$ and $f_2(1-x)$ be the rewards for choosing $G_1$ and $G_2$, given the population shares $x$ in $G_1$. The rational segment's evolutionary dynamics are driven by the payoff differential:
\begin{equation}
\Delta f(x) = f_1(x) - f_2(1-x).
\end{equation}
The functional forms of $f_1$ and $f_2$ correspond to game paradigms.

\subsection{Analysis in Three Game Paradigms}

By remaining locked in Potential Attractors, Willed Agents restrict the accessible phase space, constraining the system state $x$ to a \textbf{Feasible Region} $\Omega = [n_1, 1 - n_2]$.

Rational agents drive system dynamics by switching to higher-reward groups. We model this evolution via a Langevin equation subject to social noise $\sigma$:
\begin{equation}
dx_t = \alpha \cdot m \cdot \Delta f(x_t) dt + \sigma dW_t,
\end{equation}
where $\alpha$ is the agent movement rate and $W_t$ is a Wiener process representing choice volatility.
Natural equilibria are defined as $\mathcal{E} = \{x^* \mid \Delta f(x^*) = 0\}$, with stable equilibria $\mathcal{E}_s = \mathcal{E} \cap \{x^* \mid (\Delta f)'(x^*) < 0\}$. However, overall stability is also bounded by $\Omega$:

\begin{theorem}
A state $x^* \in \Omega$ is a stable equilibrium if it satisfies one of the following:
\begin{enumerate}
    \item \textbf{Interior Stability:} $x^* \in (n_1, 1-n_2)$ and $x^* \in \mathcal{E}_s$.
    \item \textbf{Lower Boundary Stability:} $x^* = n_1$ and $\Delta f(n_1) < 0$.
    \item \textbf{Upper Boundary Stability:} $x^* = 1 - n_2$ and $\Delta f(1 - n_2) > 0$.
\end{enumerate}
\end{theorem}

\begin{figure}[t]
    \centering
    \includegraphics[width=\linewidth]{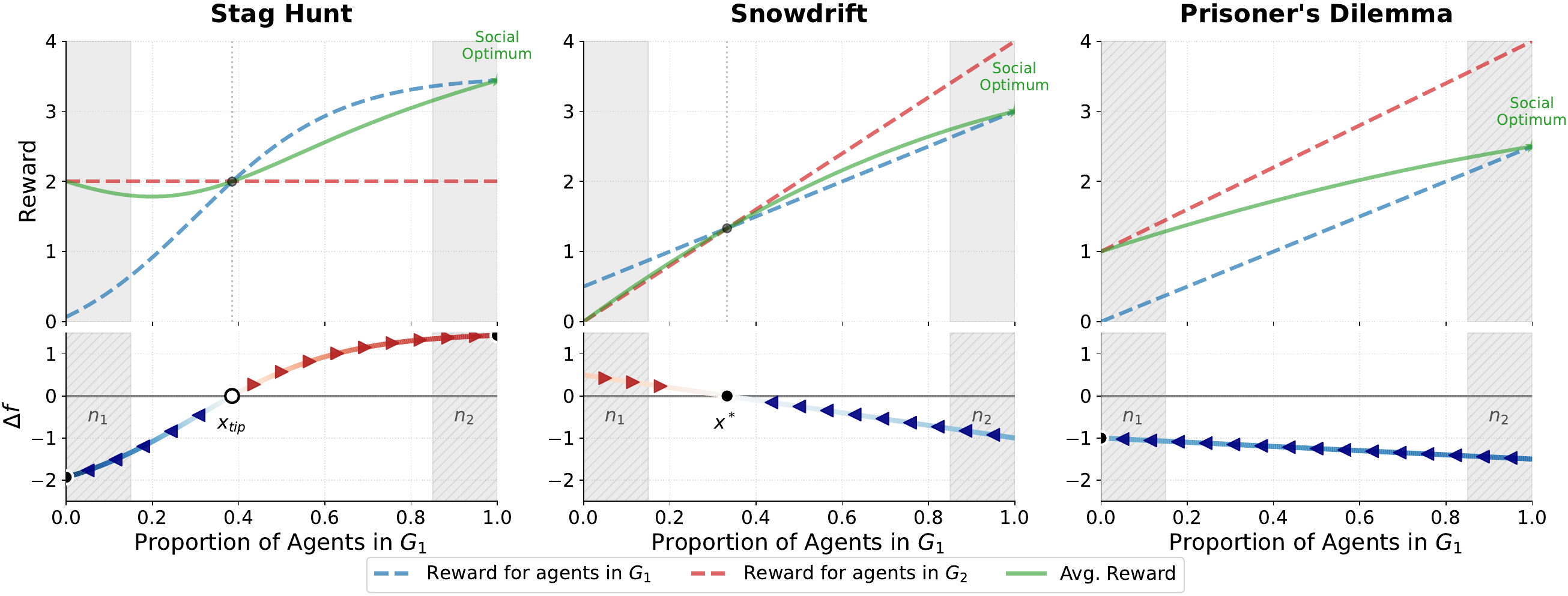}
    \caption{Willed Constraints in Canonical Games.
    \textbf{Top:} Payoff functions ($f_1, f_2$) and average reward for Stag Hunt Game, Snowdrift Game, and Prisoner's Dilemma.
    \textbf{Bottom:} The evolutionary gradient $\Delta f(x)$ and equilibrium points. 
    Hatched regions indicate the excluded state space due to Willed Agents ($n_1, n_2$), defining the feasible region $\Omega$.}
\label{fig:paradigms}
\end{figure}

We apply this framework to observe how $n_1$ and $n_2$ reshape three canonical games (Fig.~\ref{fig:paradigms}).

\subsubsection{The Stag Hunt Game}

In the Stag Hunt Game, increasing coordination yields higher rewards, creating stable equilibria at $x=0$ and $x=1$ separated by an unstable tipping point $x_{tip}$. The interaction between $\Omega$ and $x_{tip}$ yields three regimes:

\textbf{1. Degeneration into Coordination ($n_1 > x_{tip}$):} If Willed Cooperators exceed the tipping point, $x_{tip}$ is excluded from $\Omega$. Because $\Delta f(x) > 0$ across $\Omega$, the system inevitably converges to $x = 1 - n_2$, eliminating the coordination trap.

\begin{wrapfigure}[18]{r}{0pt}
    \centering
    \includegraphics[width=0.35\linewidth]{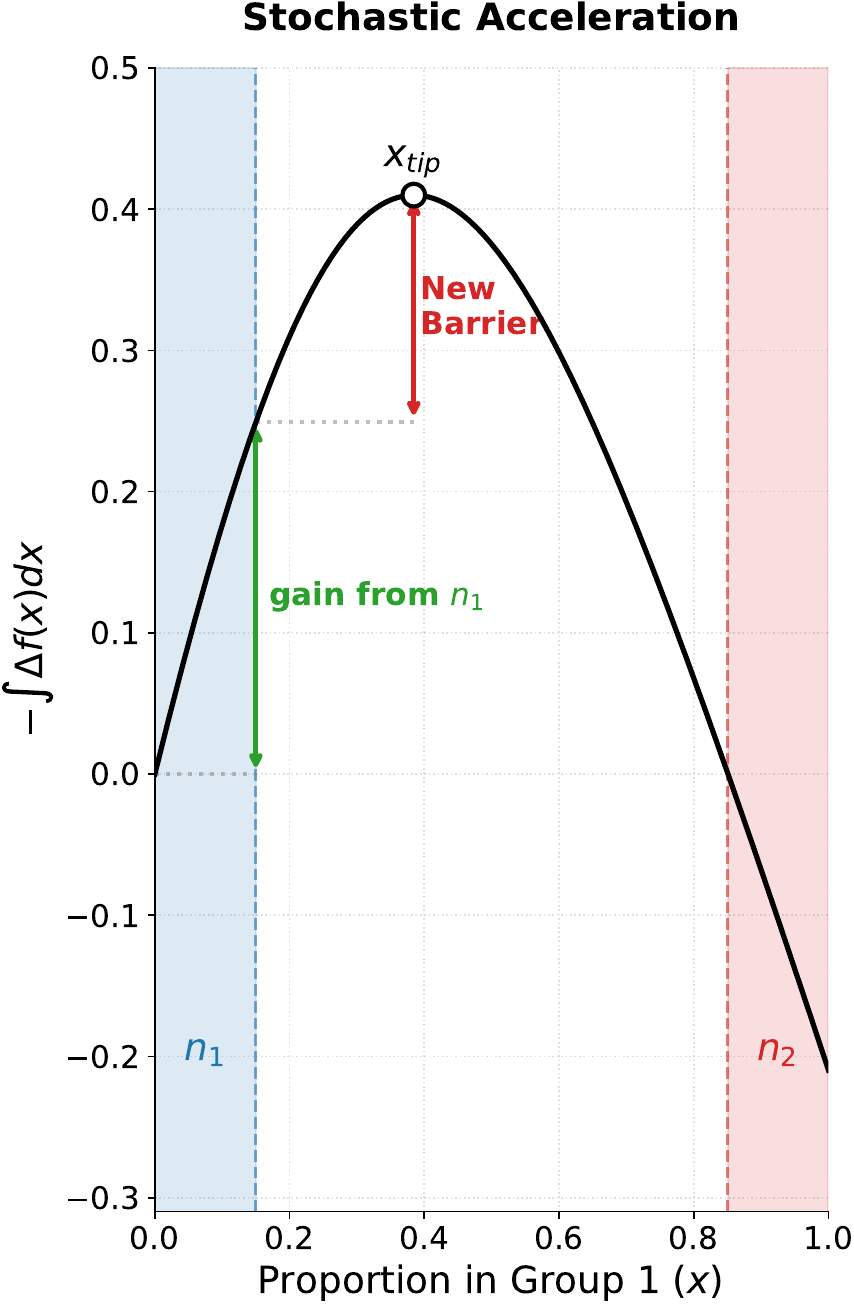}
    \caption{$n_1$ reduces the potential barrier, exponentially accelerating coordination.}
    \label{fig:stochastic_acceleration}
\end{wrapfigure}

\textbf{2. Degeneration into Dilemma ($1 - n_2 < x_{tip}$):} If Willed Defectors push the $\Omega$ ceiling below $x_{tip}$, the cooperative basin becomes inaccessible. The game functionally degenerates into a Prisoner's Dilemma, collapsing to $x = n_1$.

\textbf{3. Stochastic Acceleration ($n_1 \le x_{tip} \le 1 - n_2$):} If $x_{tip}$ remains within $\Omega$, Willed Cooperators cannot deterministically force coordination. However, raising the starting floor to $x=n_1$ shortens the distance against the gradient, increasing the probability that social noise bridges the gap to $x_{tip}$. By Kramers' Rate Law~\citep{kramers1940brownian}, the expected escape time $\tau$ drops exponentially as $n_1$ grows (Fig.~\ref{fig:stochastic_acceleration}):

\begin{equation}
    \tau \propto \exp\left( -\frac{2}{\sigma^2} \int_{n_1}^{x_{tip}} \Delta f(x) dx \right)
\end{equation}

\subsubsection{The Snowdrift Game}

In the Snowdrift Game, $\Delta f(x)$ transitions from positive to negative, establishing a stable interior equilibrium $x^*$. $\Omega$ acts as a window strictly limiting outcomes:

\textbf{1. Over-Cooperation ($n_1 > x^*$):} If Willed Cooperators exceed the optimal ratio, Rational Agents fully defect, pushing the system to $x = n_1$.

\textbf{2. Under-Cooperation ($1 - n_2 < x^*$):} If Willed Defectors restrict the ceiling below $x^*$, Rational Agents fully cooperate, and the system converges to $x = 1 - n_2$.

\textbf{3. Neutral Substitution ($n_1 \le x^* \le 1 - n_2$):} If $x^*$ lies within $\Omega$, final outcomes remain unchanged. Rational agents simply reduce their efforts in response to Willed Cooperators' fixed contributions, maintaining overall cooperation at $x^*$.

\subsubsection{The Prisoner's Dilemma}

Because defection is strictly dominant ($\Delta f(x) < 0$) in the Prisoner's Dilemma , the gradient continuously drives the system downward. The upper boundary $1-n_2$ is irrelevant, and the system collapses to the lower boundary $x^* = n_1$. Here, Rational Agents fully defect, while Willed Agents provide a stability floor and prevent total utilitarian collapse, albeit via their own exploitation.

\section{Experiments}
\subsection{Experimental Setup}
\begin{wrapfigure}[15]{r}{0pt}
    \centering
    \includegraphics[width=0.5\linewidth]{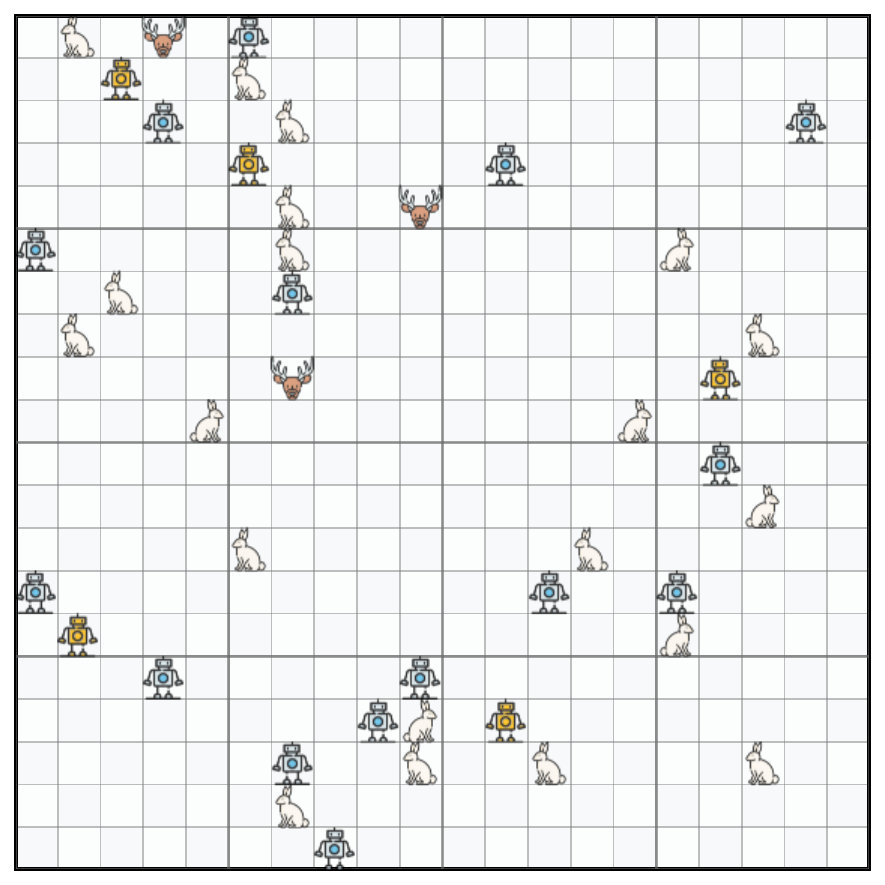}
    \caption{Overview of Markov Stag-Hunt Game.}
    \label{fig:setup}
\end{wrapfigure}
The infinite population analysis assumes that the movements of a single agent are instant and have a negligible impact on the group. However, real-world coordination occurs in finite populations with temporal constraints and spatial dynamics. In such settings, behavior is a dynamic process: individual actions are continuously observed by peers, actively shaping their belief estimations and judgments of optimal behavior over time.

To investigate the impact of Will in this context, we utilize a grid-world instantiation of the \textbf{Markov Stag Hunt Game} (Fig.~\ref{fig:setup}). We simulate an $H \times W$ grid populated by $N$ agents, $N_h$ hares, and $N_s$ stags. Each episode lasts $T$ steps. At each step, agents act synchronously by selecting from six actions: \{idle, left, right, up, down, hunt\}. Hunting a hare is a unilateral action yielding an individual reward $R_h$. Hunting a stag is a cooperative action that succeeds only if at least $\theta$ agents are co-located with the target, yielding a shared total reward $R_s$. Each agent can hunt only one prey per episode. The difficulty and incentive for cooperation are parameterized by the coordination threshold $\theta$ and the effective individual share $\bar{R_s} = R_s/\theta$. Unless otherwise specified, our experiments use $H = W = 20$, $N = 20$, $N_s = 3$, $N_h = 20$, $R_h = 1$, and $\bar{R_s} = 5$.

\subsubsection{Agent Implementation}
We instantiate the theoretical types defined in the previous section as distinct decision-making modes:

\begin{itemize}
    \item \textbf{Willed Mode:} The agent operates as a potential minimizer (Eq.~\ref{eq:potential}). The potential field is defined by the Manhattan distance $d_M$ to the target prey: $D_i(\mathbf{s}) = \inf_{g \in G_i} d_M(p_i(\mathbf{s}), p_g)$, where $p_i(\mathbf{s})$ and $p_g$ are the positions of agent $i$ and prey $g$, respectively.
    
    \item \textbf{Rational Mode:} The agent maximizes expected extrinsic reward via model-based planning. 
    First, it employs a Bayesian Theory of Mind \citep{baker2011bayesian} to infer peer goals from observed trajectories. Assuming a uniform prior $b_{ij}^0(g_j) = 1 / (N_s + N_h)$, the belief $b_{ij}^t$ over agent $j$'s target $g_j$ updates via $b_{ij}^{t+1}(g_j) \propto b_{ij}^{t}(g_j)Pr(a_j^t \mid \mathbf{s}^t, g_j)$, utilizing a Boltzmann likelihood $Pr(a_j \mid \mathbf{s}^t, g) \propto \exp(-\beta d_M(p_j(\mathbf{s}^t, a_j), p_g))$, where $\beta$ is the rationality coefficient. 
    Second, it performs $K$ Monte Carlo simulations. In each simulation $k$, peer targets are sampled from the posterior $\mathbf{g}_{-i} \sim \prod_{j \neq i} b_{ij}^{t}(g_j)$, assuming peers follow the aforementioned Boltzmann policy. 
    Finally, the agent estimates the value of hunting each prey $g$ as $V_i^k(g) = \sum_{\tau=0}^{T-t} \gamma^\tau R_i^{k, \tau}$, where $R_i^{k, \tau}$ is the reward gained at future step $\tau$. It selects the target that maximizes the average estimated reward $\sum_{k=1}^K V_i^k(g) / K$ \citep{huang2024efficient}. (Parameters used: $\beta = 10, K = 15, \gamma = 0.98$).
\end{itemize}

In complex sequential interactions, rational calculation is often computationally expensive and unreliable early in an episode due to high uncertainty and insufficient observations. Thus, agents can initially rely on the Willed Mode before switching to the Rational Mode. We formalize this transition using the \textbf{will strength} parameter $\alpha \in [-1, 1]$, which controls the temporal duration of the agent's commitment. For the first $|\alpha|T$ steps, the agent follows the Willed Mode; the sign determines the target type ($\alpha > 0$ for the Will to Stag, and $\alpha < 0$ for the Will to Hare). For the remaining steps, the agent switches to the Rational Mode. We manipulate the population \textbf{composition} (proportion of Willed Agents) and \textbf{will strength} ($\alpha$) to examine how these constraints shape collective coordination.

\subsection{Effect of Population Composition on Cooperation}
We examine how the density of Willed Agents influences collective outcomes by fixing agents to specific decision-making modes: the Will to Stag ($\alpha = 1$), the Rational ($\alpha = 0$), and the Will to Hare ($\alpha = -1$). We analyze the normalized group payoff across varying cooperation thresholds $\theta$ while keeping the incentive for cooperation constant at $\bar{R_s} = 5$.

\begin{figure}[t]
    \centering
    \includegraphics[width = \linewidth]{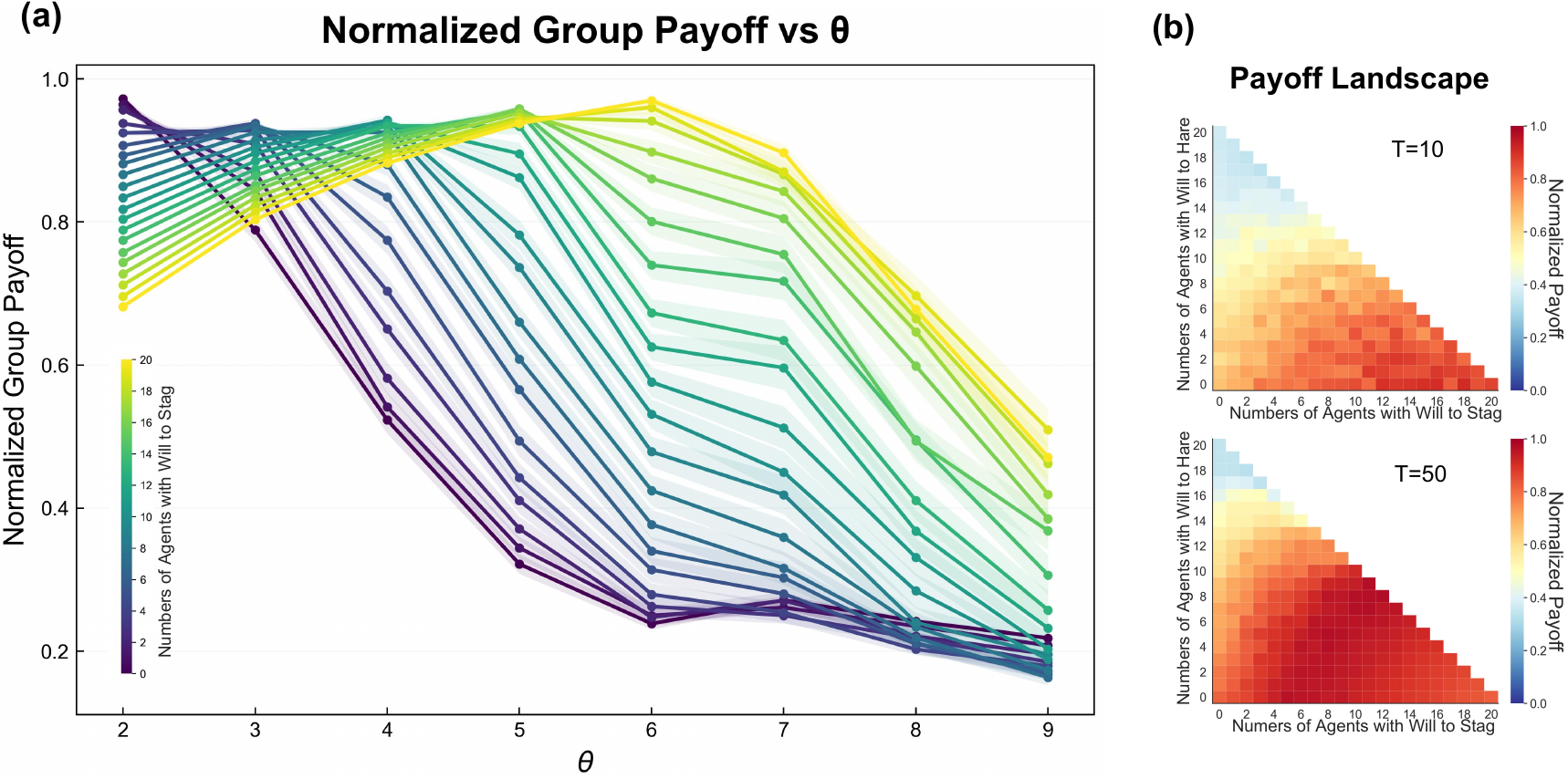}
    \caption{(a) Normalized group payoff vs. cooperation threshold $\theta$ for varying numbers of agents with Will to Stag. (b) Normalized group payoff landscape for different agent compositions at $\theta = 3$.}
    \label{fig:composition_results}
\end{figure}

\subsubsection{The Catalytic Role of Will}
Our infinite population analysis suggested that the Will to Stag benefits the group. To verify if this holds in a spatiotemporal environment, we isolate the interaction between Agents with Will to Stag and Rational Agents (Fig.~\ref{fig:composition_results}a).

Results indicate that Willed Agents function as \textit{cooperation catalysts}, though their efficacy is strictly context-dependent based on the difficulty $\theta$. When cooperation demands are minimal ($\theta = 2$), increasing the number of Willed Agents ($n_1$) actually reduces payoff, as persistence imposes unnecessary rigidity on a task that Rational Agents can easily solve. As difficulty increases, the presence of Willed Agents becomes beneficial. When $3 \leq \theta \leq 4$, this benefit is non-monotonic, where a small number improves performance but too many reduce it. When $5 \leq \theta \leq 6$, group payoff increases monotonically with $n_1$, and the gains from will are most pronounced. However, under extreme cooperation demands ($\theta \geq 7$), small additions of willed agents are insufficient; a critical mass is required to trigger any improvement in outcomes. 

Collectively, these results show that \textbf{while will can enhance cooperation, it exhibits both critical mass effects and non-monotonicities depending on cooperation difficulty}.

\subsubsection{Full Composition Analysis under Temporal Constraints}
We further analyze the full ternary composition space under varying time horizons ($T=10$ and $T=50$) with $\theta=3$. As shown in Fig.~\ref{fig:composition_results}b, shorter episodes ($T = 10$) require a substantially higher proportion of agents with Will to Stag to achieve high performance compared to longer episodes.

This result aligns with the analysis of the Stag Hunt Game in infinite population. Rational Agents require time to adjust beliefs. Given sufficient steps, Rational Agents can observe trajectories, infer peer intent, and align with the Willed Agents. However, under tight temporal constraints, the system relies on a critical density of the Will to Stag to "fast-track" the group toward the Stag equilibrium before the episode terminates. 

Overall, \textbf{Willed Agents provide the initial momentum to overcome cooperation friction, which is crucial when the cooperation threshold is high or time is scarce}.

\subsection{Effect of Will Strength on Cooperation}

While the previous section established that the presence of Willed Agents catalyzes coordination, those agents operated with fixed, infinite persistence. We now investigate the impact of \textbf{will strength} ($\alpha \in [-1, 1]$), which governs the duration of an agent's commitment before reverting to rational mode. This allows us to examine how the intensity of individual ``will'' translates into collective outcomes.

\subsubsection{Homogeneous Populations}

\begin{wrapfigure}[18]{r}{0pt}
    \centering
    \includegraphics[width=0.6\linewidth]{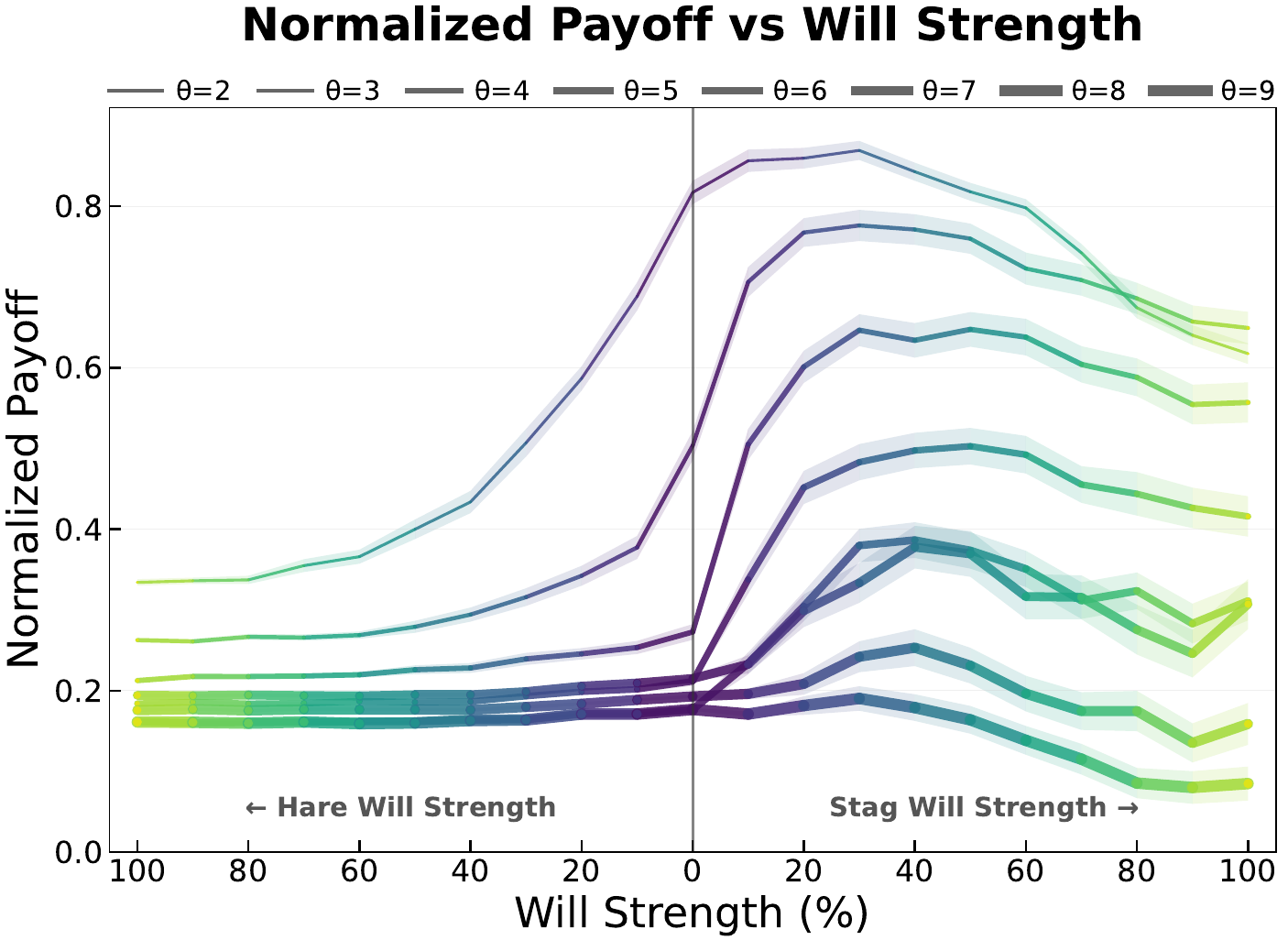}
    \caption{Normalized group payoff as a function of will strength $\alpha$ across coordination thresholds $\theta$. $\bar{R_s} = 5, T = 10$. Shaded regions indicate 95\% confidence intervals over 300 independent episodes.}
    \label{fig:strength_results}
\end{wrapfigure}

We first analyze populations where all agents share a uniform will strength $\alpha$ under tight temporal constraints ($T=10$). As shown in Fig.~\ref{fig:strength_results}, the relationship between will strength and group payoff is non-linear and strictly modulated by the coordination difficulty $\theta$.

For the Will to Stag ($\alpha > 0$), we observe a distinct \textbf{inverted-U relationship} between will strength and performance at intermediate difficulties ($3 \leq \theta \leq 7$). A moderate level of will ($\alpha \approx 0.5$) effectively anchors agents to the cooperative equilibrium, overcoming initial coordination friction without inducing pathological stubbornness. However, as $\alpha$ approaches 1.0, the benefits diminish; excessive will induces functional rigidity, preventing agents from making necessary local adjustments,  especially when the target prey is likely to
be captured by other agents.

At the extremes of coordination difficulty, the benefits of will vanish. When coordination is trivial ($\theta = 2$), rigidity becomes a liability, and rational agents can coordinate spontaneously and adjust your goals promptly based on the observation. and rational population ($\alpha=0$) outperform population with $\alpha > 0.4$. Conversely, when coordination is severe ($\theta \ge 8$), the high requirement for simultaneous presence makes stag hunting statistically improbable. In these cases, the inflexible pursuit of stags yields lower returns than the rational baseline of safe hare-hunting. Finally, the Will to Hare ($\alpha < 0$) consistently underperforms or matches the baseline, confirming that persistence is evolutionarily advantageous only when directed toward high-risk, high-reward equilibria that rational agents fail to reach.

\subsubsection{Heterogeneous Populations}

Biological and social systems are rarely homogeneous. To determine the evolutionarily stable distribution of will, we utilize a genetic algorithm to optimize the population's composition of $\alpha$ ($N=10, N_s = 2, N_h = 10, \bar{R_s} = 10, T = 10$).
 
\begin{figure}[tb]
    \centering
    \includegraphics[width = \linewidth]{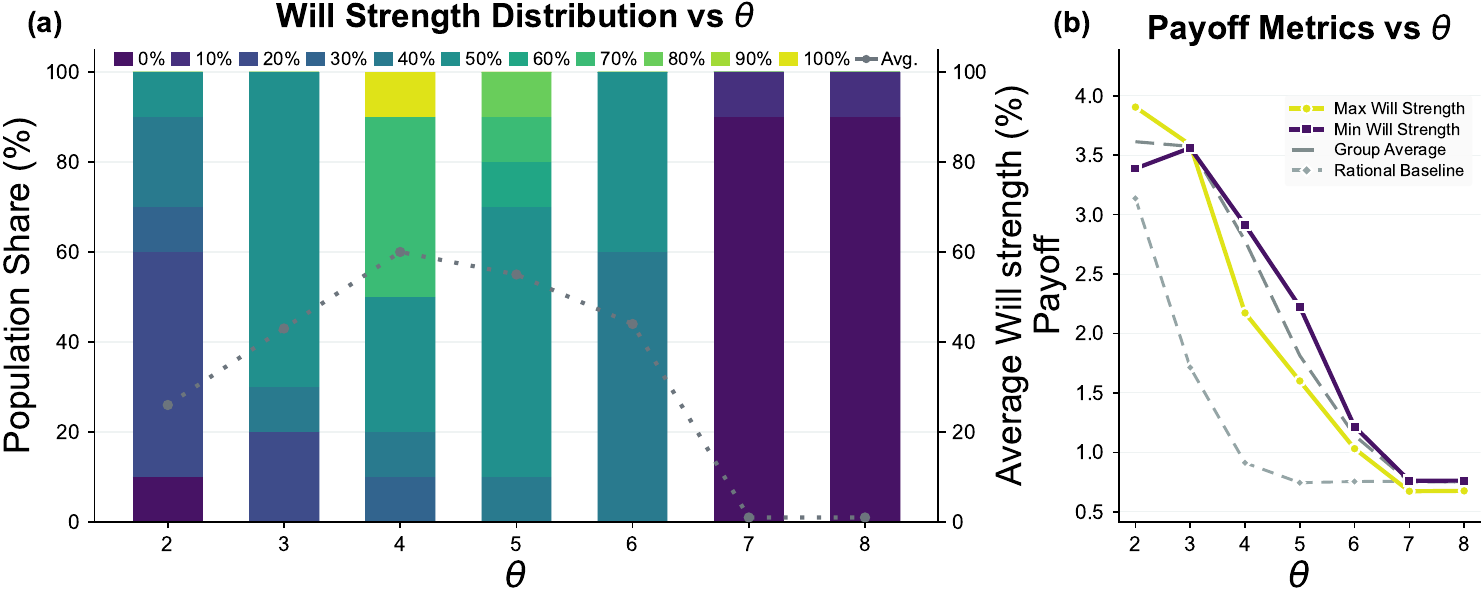}
    \caption{(a) Optimal will strength distribution across $\theta$. Stacked bars indicate the population share at each strength level. The dashed line tracks the mean. (b) Individual payoff for maximal vs. minimal will strength, with dashed lines indicating optimized group and rational baseline payoffs.}
    \label{fig:hetero_results}
\end{figure}

The results (Fig.~\ref{fig:hetero_results}a) reveal that the ideal group composition varies systematically with coordination difficulty. The optimal average will strength peaks at $\theta = 4$ with average $\alpha \approx 0.6$ before declining. 
When coordination is feasible, the optimal population exhibits high heterogeneity: high-$\alpha$ agents anchor the cooperative goal, creating a focal point, while lower-$\alpha$ agents retain the flexibility to adapt to stochastic dynamics. As $\theta$ increases beyond critical thresholds, the population converges toward homogeneity and rationality, abandoning the high-risk stag strategy.

Fig.~\ref{fig:hetero_results}b highlights the cost of this specialization. At most coordination thresholds, agents with high will strength earn lower individual payoffs than their lower-will counterparts. High-will agents lower the energetic barrier for coordination but pay a fitness cost for their rigidity. This suggests that the Will to Stag functions as a form of altruistic commitment, enabling the group to achieve superior outcomes compared to a purely rational population, often at the expense of the willed individual's local utility.

\subsection{Endogenous Will Selection}

In previous experiments, target states were exogenously assigned. A fully autonomous agent, however, must internally generate the ``will'' to pursue a specific goal. We implement hybrid agents that utilize Monte Carlo Planning (Rational Mode) to identify a high-value target, then switch to Potential Minimization (Willed Mode) to greedily pursue it without re-planning. Let $k \in [1/T, 1]$ be the \textit{Rational Ratio}, representing the proportion of timesteps allocated to re-evaluation. We compare two temporal integration strategies:

\begin{enumerate}
    \item \textbf{Intermittent:} The agent engages in rational planning periodically (every $1/k$ steps) to update its target. In the interim steps, it acts as a Willed Agent committed to the most recently selected target. This mimics an agent that frequently "checks in" on the validity of its goal.
    \item \textbf{Phased:} The agent confines rational planning to the initial phase of the episode (the first $kT$ steps). Once this phase concludes, the agent "locks in" the final target and operates in Willed Mode for the remainder of the episode.
\end{enumerate}

\begin{table}[tb]
\caption{Average normalized group payoff under endogenous goal selection ($\theta=4$). "Instant" refers to an agent that plans only at $t=0$ and maintains that commitment for the entire episode. Data averaged over 300 independent episodes.}
\centering
\renewcommand{\arraystretch}{0.85}
\setlength{\tabcolsep}{4pt}
\begin{tabular}{lccc}
\toprule
Strategy & Rational Ratio & $\bar{R_s} = 10$ & $\bar{R_s} = 50$ \\
\midrule
Pure Rational & 100\% & \textbf{0.607}\err{0.024} & 0.716\err{0.027} \\
\midrule
Intermittent & 50\% & 0.577\err{0.023} & 0.700\err{0.027} \\
Phased & 50\% & 0.587\err{0.024} & 0.693\err{0.025} \\
\midrule
Intermittent & 20\% & 0.521\err{0.022} & 0.687\err{0.025} \\
Phased & 20\% & 0.578\err{0.026} & 0.678\err{0.026} \\
\midrule
Intermittent & 10\% & 0.505\err{0.019} & 0.724\err{0.025} \\
Phased & 10\% & 0.562\err{0.023} & 0.684\err{0.025} \\
\midrule
Instant& 2\% & 0.510\err{0.019} & \textbf{0.782}\err{0.024} \\
\bottomrule
\end{tabular}
\label{table:endogenous_will}
\end{table}

We evaluate these strategies under coordination difficulty $\theta=4$ (Tab.~\ref{table:endogenous_will}). The results highlight a trade-off between flexibility and commitment. At equivalent rational ratios, the Phased strategy excels when cooperative rewards are low ($\bar{R_s} = 10$), as early commitment prevents abandoning marginal opportunities. Conversely, the Intermittent strategy performs better at high cooperative rewards, where false-positive commitments are costlier.

Crucially, under high incentives ($\bar{R_s} = 50$), the Instant strategy (a single initial plan followed by unwavering execution) significantly outperforms the Pure Rational baseline. This challenges the assumption that more information processing yields better outcomes, as continuous re-evaluation makes rational agents hypersensitive to stochastic noise, often leading to the premature abandonment of joint goals. 

While suspending rational planning undeniably serves as a computational cost-saving shortcut, its primary evolutionary and social function is far richer: \textbf{it actively signals predictability to peers.} By intentionally constraining its own cognitive flexibility and ``binding its will'' to a target, the agent transforms into a reliable focal point, effectively solving the coordination problem. This empirically validates the concept of \textit{Resolute Choice}~\citep{mcclennen1990rationality}: under high-stakes uncertainty, the optimal strategy is often to decide wisely once, and then refuse to reconsider.

\section{Conclusion and Discussion}

We have proposed a decision-making framework where social agents are driven by potential minimization rather than pure utility maximization. By constraining the feasible solution space, willed agents act as cooperation catalysts, enabling populations to traverse high coordination thresholds that purely rational groups fail to overcome.

Although ``willed'' behavior can positively impact society, its local fitness costs raise questions about its evolutionary survival. This persistence may be explained by cooperative mechanisms in evolutionary game theory~\citep{nowak2006five, santos2006evolutionary} and economic risk models, where some agents pursue high-risk, high-reward goals~\citep{kahneman2013prospect, schildberg2018risk}.

Future research should explicitly investigate the evolutionary dynamics of willed agents across diverse spatiotemporal conditions. Additionally, integrating endogenous Will Selection with multi-agent reinforcement learning could explore how agents learn a meta-policy, dynamically regulating \textit{when} to bind their will to balance the stability required for coordination against the flexibility needed for adaptation.

\section*{Acknowledgement}
This work is supported by the National Science and Technology Major Project (No. 2022ZD0114900). This work is also supported by Beijing Natural Science Foundation (No. 4264117).

\printbibliography

\end{document}

%% file: sections/introduction_3.tex
\section{Introduction}
Social dilemmas arise when individually rational decision-making systematically produces collectively suboptimal outcomes \citep{olson1971logic, hardin1968tragedy, schelling2006micromotives}. Across classical economics and cognitive science, agents are typically modeled as efficient optimizers, such as utility maximizers in game theory and Bayesian planners in computational neuroscience, who continuously adjust behavior to maximize expected reward \citep{simon1956rational, camerer2003behavioral}. In social dilemmas, this commitment to scalar maximization takes the form of repeated local cost--benefit evaluation under uncertainty \citep{camerer2004cognitive, van1990tacit}. A substantial body of work shows that such optimizing dynamics can amplify perceived risk and strategic caution, driving agents toward safe but collectively inferior equilibria, as in the Stag Hunt game, even when mutually beneficial coordination is feasible~\citep{arthur1994inductive, hommes2013behavioral}.

Most formal accounts of cooperation failures approach the problem by altering agents’ incentives or the structure of their interactions. One class of approaches promotes cooperation by reshaping local payoffs or preferences, including punishment mechanisms \citep{fehr2000cooperation, koster2022spurious}, reputation systems \citep{nowak2005evolution, michel2024evolution}, and social preference models \citep{rabin1992incorporating, fehr1999theory, bogaert2008social, hughes2018inequity, kong2024learning}. A second class stabilizes cooperation through institutional rules, governance structures, or centralized enforcement \citep{greif2006institutions, ostrom2009understanding, acemoglu2015nations}. While effective in structured settings, institutional and governance-based approaches rely on externally imposed rules or centralized enforcement, which incur ongoing implementation and monitoring costs \citep{north1990institutions, tirole2017economics}.

Despite their differences, these approaches share a common premise: cooperation fails because agents compute or incentivize too little, and can therefore be remedied by additive computation. This perspective leaves largely unexplored the possibility that the problem lies not in insufficient optimization, but in the very reliance on continuous rational computation itself~\citep{colman2003cooperation, mcclennen1990rationality}. Accordingly, this paper asks whether a cognitive mechanism beyond the rational computation paradigm can enable groups to escape suboptimal non-cooperative equilibria without altering the underlying payoff structure.

To address this question, we introduce the concept of \emph{will}. Rooted in political philosophy and social theory, \emph{will} is a way of explaining how intentions, commitments, and collective expectations persist over time. At the individual level, \emph{will} refers to an agent’s capacity to form and sustain intentions beyond the moment of choice, as emphasized in philosophical accounts of intention and agency \citep{frankfurt2018freedom, bratman1987intention}. This capacity resonates with psychological research on implementation intentions and action control, which demonstrates that individuals who form specific plans exhibit reduced deliberation and increased persistence in goal pursuit, even when immediate incentives favor alternative actions~\citep{gollwitzer1999implementation, heckhausen2018motivation,brandstatter2001implementation}. Such findings suggest that the deliberate suspension of ongoing cost--benefit evaluation is not merely a philosophical construct but an empirically documented feature of human cognition. At the social level, it describes how interdependent agents come to rely on one another’s actions, allowing mutual expectations to stabilize and coordinated order to emerge \citep{hobbes2004leviathan, rousseau1997rousseau}. Building on this dual perspective, we conceptualize \emph{will} as a cognitive mechanism. 
Once a target is selected, a willed agent persistently pursues states that bring it closer to the target over time, regardless of cost-benefit fluctuation.
Notably, \textit{will} is not an alternative decision rule to classical models such as expected utility maximization, bounded rationality, or dual-process frameworks, but a complementary mechanism that operates on top of them.

Although decision research offers several constructs related to \emph{will}, they remain fundamentally grounded in local cost--benefit evaluation. They promote behavioral persistence by anchoring goals, value signals, or choice sets, while leaving ongoing action continuously subject to reassessment as circumstances change. Commitment, for example, constrains future behavior by restricting the set of admissible choices in advance \citep{Elster1984-ELSUAT}, yet the agent continues to evaluate which action to select among the remaining options at each decision point. Goal or intention stability maintains fixed goal representations over time \citep{georgeff1998belief, braver2012variable}, but the means of pursuing those goals remain open to ongoing cost--benefit recalculation. Motivation sustains the strength of value or reward signals \citep{kanfer1990motivation, ryan2000intrinsic}, yet the decision process still responds to momentary fluctuations in perceived costs and benefits. Persistence or grit operates at a descriptive level without specifying an underlying decision mechanism, characterizing sustained effort toward long-term goals despite obstacles \citep{duckworth2014self}. Crucially, deliberately suspending local recalculation is a documented cognitive strategy. Bounded rationality research shows humans strategically limit ongoing computation to reduce deliberative overhead and signal predictability to partners \citep{simon1956rational}. This temporary commitment must be distinguished from zealot models that fix strategies irrespective of environmental structure \citep{masuda2012evolution, cardillo2019critical}, ``cooperate without looking'' that precludes strategic responsiveness \citep{hoffman2015cooperate}, Bayesian Theory of Mind that relies on continuous mental-state inference \citep{peysakhovich2018prosocial, shum2019theory, kleimanweiner2025evolving}, and virtual bargaining that coordinate behavior through mutual representations of joint plans \citep{colman2003cooperation}. By contrast, \emph{will} interrupts local evaluation: rather than recalculating at every step, a willed agent maintains a fixed course of action as circumstances unfold.

To the best of our knowledge, this paper provides the first mathematical formulation of \textit{will} for multi-agent decision-making.
Based on this formulation, the paper develops three lines of analysis.
First, we provide a formal mathematical formulation of \emph{will}, centered on the property that willed agents consistently pursue target states. 
Second, we analyze the interaction dynamics of an infinite population across three canonical social dilemma paradigms (Stag Hunt game, Snowdrift game, and Prisoner's Dilemma), examining how \textit{will} reshapes individual behavior and collective outcomes. 
Third, in a more complex spatiotemporal social dilemma, the Markov Stag Hunt game, the impact of \emph{will} on cooperation is investigated in terms of the proportion of Willed Agents and the distribution of will strength.
Results show that Willed Agents function as "cooperation catalysts", enabling groups to surmount high-risk thresholds where purely utility maximization fails, and that optimal performance relies on heterogeneous will strength within the population.
Moreover, a parsimonious experiment for the Willed Agents with autonomously target generation shows that autonomously suspending rational re-evaluation can significantly outperform continuous optimizers. 
Overall, while \textit{will} forgoes continuous reward optimization, it functions as a critical mechanism for stabilizing cooperation in social dilemmas, where agents face conflicting incentives and strategic uncertainty.